\documentclass[twocolumn,showkeys,showpacs,preprintnumbers,prd,superscriptaddress,nofootinbib]{revtex4}

\usepackage{graphicx}
\usepackage{epsf}
\usepackage{bm}
\usepackage{amsmath}
\usepackage{amsfonts}
\usepackage{amssymb}
\usepackage{epstopdf}
\usepackage{natbib}
\usepackage{hyperref}
\usepackage{color}
\usepackage{verbatim}
\usepackage{multirow}
\usepackage{bm}

\begin{document}

\title{Searching for modified gravity in the astrophysical gravitational wave background: Application to ground-based interferometers}

\author{Rafael C. Nunes}
\email{rafadcnunes@gmail.com}
\affiliation{Divis\~ao de Astrof\'isica, Instituto Nacional de Pesquisas Espaciais, Avenida dos Astronautas 1758, S\~ao Jos\'e dos Campos, 12227-010, S\~ao Paulo, Brazil}

\begin{abstract}
We investigate how the propagation of an astrophysical gravitational wave background (AGWB) is modified over cosmological volumes when considering theories beyond general relativity of the type Horndeski gravity. We first deduce an amplitude correction on the AGWB induced for the presence of a possible running in the Planck mass. Then, we apply the spectral noise density from some ground-based interferometers, namely, the Advanced LIGO (aLIGO), Einstein Telescope (ET) and Cosmic Explore (CE), to evaluate the signal-to-noise ratio (SNR) as a function of the amplitude of the running of the Planck mass for two different scenarios. We find that for observation time period $\gtrsim$ 5 yrs and  $\gtrsim$ 1 yr, we can have a significant signal of the AGWB in the band [1-100] Hz from the ET and CE sensitivity, respectively. Using Fisher information, we find some forecast bounds, and we deduce $\lesssim$ 27\% and $\lesssim$ 18\% correction at 1$\sigma$ confidence level on the amplitude of the running of the Planck mass from ET and CE, respectively. It is clear that a detection of a AGWB in future can open a new window to probe the nature of gravity with good accuracy.
\end{abstract}

\maketitle

\section{Introduction}

One century after the formulation of Einstein’s general relativity (GR), the gravitational waves (GWs), being one of the main theoretical predictions of GR, are recently observed in LIGO/VIRGO \cite{ligo01}. 
It was possible to observe the Universe, and discover several objects and physical phenomena in the last 100 years through various observations mainly via electromagnetic signal. Now, discovery of the GWs, has opened a new observational window to investigate the Universe under a new spectrum of possibilities. Also, through possible joint observation in GWs and electromagnetic signal, called the multi-messenger search, just like the recent GW170817 and GRB 170817A events \cite{Gw07,Gw08}, and other events should be detected soon.

The GWs are emitted mainly by individual binary systems, like for instance, from binary black holes (BBH), binary neutron stars (BNS) and binary black hole-neutron stars (BBH-NS). It is expected that the superposition of the signal from these sources over cosmological volumes can form an astrophysical gravitational waves background (AGWB) \cite{Phinney,A1,A2,A3,R1,R2,R3,R4}. The AGWB is characterized by the energy density parameter $\Omega_{\rm GW}(f)$, which represents the present-day fractional energy as a function of frequency $f$. The AGWB signal strongly depends on the type of sources that produce them, and we expect that signal to exist in the most diverse frequencies \cite{AGWB1,AGWB2,AGWB3,AGWB4,AGWB5,AGWB6,AGWB7,AGWB8,AGWB9,AGWB10,AGWB11,AGWB12,AGWB13,AGWB14}. Until the present moment, the AGWB have not been detected, and only some upper limits on the stochastic gravitational-wave background signal have been obtained. The LIGO/VIRGO collaboration reported an astrophysical background with amplitude $< 4.8 \times 10^{-8}$ \cite{LIGO_1} and 
$\Omega_{\rm GW} = 1.8^{+2.7}_{-1.3} \times 10^{-9}$ at 25 Hz in \cite{LIGO_2}. With the improvements in instrumental sensitivity in the coming years, as well as from the prospects of the future detectors like the Einstein Telescope (ET) \cite{ET}, Cosmic Explorer (CE) \cite{CE} and LISA \cite{LISA_1,LISA_2}, it is expected to achieve enough sensitivity to detect the AGWB. 

There are theoretical and observational reasons to believe that GR should be modified when gravitational fields are strong and/or on large scales. From an observational point of view, the physical mechanism responsible for accelerating the Universe at late times is still an open question, and new degrees of freedom of the gravitational origin are alternatives to explain such an accelerated stage (see \cite{DE_review,MG_review01,MG_review03} for review). Theories beyond GR can serve as alternatives to explain the current tension in the Hubble constant that persists in the framework of the $\Lambda$CDM model \cite{MG_H0_1,MG_H0_2,MG_H0_3,MG_H0_4}. Also, modified gravity models are motivated to drive the accelerating expansion of the Universe at early times (inflation). See \cite{Berti} and references therein for motivation of modified gravity scenarios under the regime of strong gravitational field.

All the first GWs observations to date have revealed to be consistent with GR theory \cite{LIGO_MG1,LIGO_MG2}, and imposed strong bounds on modified gravity/dark energy scenarios in the local Universe \cite{GW_DE1,GW_DE2,GW_DE3}. On the other hand, going beyond general relativity means changing the properties of GWs in different ways, such as corrections in amplitude, phase, extra polarization states, etc. It is expected that on large scales and cosmological distances, the GR theory needs to be corrected and, in return, on such scales, it will change the GWs behavior/properties \cite{GW_C1,GW_C2,GW_C3,GW_C4,GW_C5,GW_C5_2,GW_C6}. Current detectors are not sensitive enough to probe the Universe at cosmological distances, not more than $z \lesssim 1$. But, certainly, some promising projects like ET, CE and LISA will be able to detect GWs events with great accuracy at cosmological distances and provide information for powerful cosmological tests \cite{GW_C7,GW_C8,GW_C9,GW_C10,GW_C11,GW_C12,GW_C13,GW_C14,GW_C15,GW_C16,GW_C17,GW_C18,GW_C19,GW_C20,GW_C21,GW_C22,GW_C23,GW_C24}.

The goal of this article is to extend the standard calculation of the gravitational wave background from compact binary coalescences. Where we consider a sum on the contributions from the binary neutron stars + black hole-neutron + black holes, incorporating corrections on the propagation of the AGWB signal over cosmic time in the presence of possible changes in GR theory. We find that the gravitational coupling, quantified in terms of the running of the Planck mass $\alpha_M$, can induce amplitude corrections on $\Omega_{\rm GW}(f)$ propagation. Then, we analyze signal-to-noise ratio as a function of $\alpha_M$ for two scenarios, assuming the possibilities $\alpha_M < 0$ and $\alpha_M > 0$, from the perspective of the LIGO, ET and CE sensitivity. We note that from ET and CE, we can have a significant signal, and then we perform a forecast analysis on the free parameter that determines the $\alpha_M$ amplitude.

The paper is structured as follows. In Section \ref{MG_AGWB}, we present the theoretical framework for the AGWB propagation in modified gravity. In Section \ref{results}, we present our main results. Lastly, in section \ref{final}, we outline our final considerations and perspectives. 

\section{Modified gravitational wave background from compact binary coalescences}
\label{MG_AGWB}

In this section, we summarize the formalism used to calculate the GWs energy spectrum $\Omega_{\rm GW}(f)$ as presented in \cite{Phinney,AGWB1}. Let us check the theories beyond the GR inducing corrections on the $\Omega_{\rm GW}(f)$ propagation, and when evaluated at the present moment, i.e, $z = 0$, we can compare the spectrum in possible GWs experiments/observatories. 

The GWs spectrum can be computed by

\begin{equation}
\label{GW_1}
\Omega_{\rm GW}(f) = \frac{1}{\rho_c} \int_{0}^{z_{\rm max}} \frac{N(z)}{1+z} \Big(\frac{dE_{\rm GW}}{d \ln f_r} \Big)  dz,
\end{equation}
where $N(z)$ is the spatial number density of GW events at $z$. The factor $(1+z)$ accounts for redshifting of GW energy since emission, and $f_r = f (1 + z)$ is the GW frequency in the source frame. The function $dE_{GW}/d \ln f_r$ quantifies the single source energy spectrum. It is convenient to replace $N(z)$ with the differential GWs event rate

\begin{equation}
\frac{dN}{dz} = N(z) 4 \pi r^2 = \frac{R(z)}{1+z} \frac{dV}{dz},
\end{equation}
where $dV/dz = 4 \pi c r^2/H(z)$ is the comoving volume element, and $r$ the comoving distance. Here, $R(z) = r_0 e(z)$ is the rate density
measured in cosmic time local to the event \cite{R_1,R_2}, where the parameter $r_0$ is the local rate density, used to estimate detection rates for different detectors. The amount $e(z)$ is a dimensionless factor which models the source rate evolution over cosmic time. The factor $(1 + z)$ in the above equation converts $R(z)$ to an earth time based quantity. Then, eq. (\ref{GW_1}) can be written as

\begin{equation}
\label{GW_2}
\Omega_{\rm GW}(f) = \frac{f r_0}{\rho_c H_0} \int_{0}^{z_{\rm max}} \frac{e(z)}{(1+z) E(z)} \Big(\frac{dE_{\rm GW}}{d \ln f} \Big) dz
\end{equation}

The GW energy spectrum for an inspiralling circular binary of component masses $m_1$ and $m_2$ is given by \cite{Thorne} 

\begin{equation}
\frac{dE_{\rm GW}}{d \ln f_r} = \frac{(\pi G)^{2/3} M_c^{5/3}}{3} f_r^{-1/3},
\end{equation}
where $M_c = M \eta^{5/3}$ is the chirp mass, $M = m_1 + m_2$ being the total mass and $\eta = m_1m_2/M^2$ the symmetric mass ratio. Inserting this into eq. (\ref{GW_2}), we have

\begin{equation}
\label{GW_3}
\Omega_{\rm GW}(f) = A \times J \times f^{2/3}, 
\end{equation}
where we have defined the quantity

\begin{equation}
A = \frac{8 r_0}{9 c^2 H_0^3} (\pi G M_c)^{5/3}
\end{equation}
and


\begin{equation}
J = \int_{0}^{z_{\rm max}} \frac{e(z) (1+z)^{-4/3}}{\sqrt{\Omega_m(1+z)^3 + (1-\Omega_m)}}.
\end{equation}

To determine the applicable frequency range of the above relation, one has $f_{min}$ well below 1 Hz, and $f_{max}$ given by the frequency at the last stable orbit during inspiral, $f_{max} = 1/(63/2^2\pi M_z)$, with $M_z = (1 + z)M$.

We are interested in checking how alternative scenarios to GR can change $\Omega_{\rm GW}(f)$. In principle, we have two major possibilities to look into this point, which can globally affect the AGWB. 
\\

i) A common feature in almost all the gravity theories beyond the GR, at level of the Universe on large scales, is that the new degrees of freedom in each theory modify the gravitational force/interaction on cosmological scales, mainly motivated to explain the late-time acceleration of the Universe (dark-energy-dominated era). This case is generally featured by an effective time variable gravitational constant, that will affect the propagation of the GWs along the cosmic expansion. See \cite{Ezquiaga_Zumalacarregui} for a review.
\\

ii) By changing gravity, we also change the generation mechanism of the gravitational radiation emitted by the binary systems. Such methodology can be quantified through the parameterized post-Einsteinian framework \cite{PPE_1,PPE_2,PPE_3,PPE_4,PPE_5}. In this case, changes in GR will modify the waveform, but keep propagation corrections on GWs unchanged. The AGWB is recently studied in the parameterized post-Einsteinian context in \cite{AGWB_PPE}.
\\

In general, at local level, the GWs information from isolated or binary systems  in strong space-time curvature regime can provide several tests to GR \cite{Berti}. But, we are interested in the AGWB, which is a global source. Thus, these two points above should be the main sources of corrections in this sense. In this work, let us focus on the case i), where the presence of some scalar field can significantly modify the gravitational force and the effective gravitational couplings vary in time at the cosmological scales, so that it is possible to see the variation of the gravitational couplings as a function of the cosmic time. As the AGWB is evaluated over large cosmic time intervals, corrections in this sense can become an interesting source of the information about gravity. It is important to note that this framework is quantifying directly dark energy effects and its fingerprint on the AGWB propagation over cosmological time. Here, these dark energy effects are not considered/significant on the compact objects structure.

As this proposal, let us formulate these corrections on the AGWB in the context of the Horndeski gravity \cite{Horndeski,Deffayet,Kobayashi11}, which is the most general scalar-tensor theory with second-order equations in $D = 4$. In appendix A, we briefly review this gravity scenario. Following \cite{GW_C3,GW_C4}, the effective Newton constant can be written as 

\begin{equation}
\label{G_eff}
\frac{G_{\rm gw}}{G_N} = \frac{M^2_{*}(0)}{M^2_{*}(z)},
\end{equation}
where $G_N$ is the Newton gravitational constant, and we define $G_{gw}$ as the gravitational coupling for GWs, where 

\begin{equation}
\label{alpha_M}
M^2_{*} = 2 G_4 \,\,\,\,\,\,\,\,  {\rm and} \,\,\,\,\,\,\,\, \alpha_M = \frac{1}{H} \frac{d \log M^2_{*}}{dt},
\end{equation}
with $M^2_{*}$ being the effective Planck mass and $\alpha_M$ the running of the Planck mass.

Interpreting the gravitational constant in eq. (\ref{GW_3}) as the gravitational coupling for GWs, we can write the spectrum as

\begin{equation}
\label{GW_4}
\Omega^{\rm MG}_{GW}(f) = \bar{A}^{-10/3}\Omega^{\rm GR}_{GW}(f),
\end{equation}
with

\begin{equation}
\label{GW_5}
\bar{A} = \exp \Big[-\frac{1}{2} \int_0^z \frac{\alpha_M(z')}{1+z'} dz' \Big],
\end{equation}
where the indices GR and MG refer to the spectrum predicted in general relativity and modified gravity, respectively. Note that this correction comes  due to the energy spectrum from the inspiralling binary systems, and can be interpreted as an amplitude correction, a cumulative effect throughout the propagation of the AGWB through the cosmic evolution.

Now, in order to move on, it is usual to choose phenomenological functional forms for the functions $\alpha_{M}$ (see, e.g., \cite{alphai_01,alphai_02}). In the present work, we will adopt two parametrizations:
\\

\textbf{Scenario I}: The sub-Horndeski gravity called by {\it No Slip Gravity}, proposed in \cite{alphai_03}. The main characteristics of this model read as the speed of gravitational wave propagation equal to the speed of light, and equality between the effective gravitational coupling strengths to matter and light, but yet different from Newton's constant, capable of generating an effective $G_{\rm gw}$. In this scenario we have

\begin{equation}
\label{Linder_model1}
\alpha_{M} = \frac{c_M (a/a_t)^{\tau}}{ [(a/a_t)^{\tau}+1]^2},
\end{equation}
where $a$ is the cosmic scale factor. The main parameter here is $c_M$,
featuring the amplitude of the running of the Planck mass. The stability condition requires $c_M > 0$ and $ 0 < \tau < 3/2$. In what follows, through all the simulations carried out in this work, we will adopt $a_t = 0.5$ and $\tau = 1$. See \cite{alphai_03} for more details.
\\

\textbf{Scenario II}: To quantify $\alpha_{M} < 0$ effects, let us consider $\alpha_{M} = c_M a^n $. Following \cite{alphai_02},  the stability conditions for $\alpha_{M} < 0$ can be summarized as: stable for $c_M < 0$ and $n > 3/2$. Throughout our results below, we will assume $n = 1$.
\\

Once $\alpha_{M}$ is defined, we can evaluate correction on the spectrum due to the modified propagation and compare the theoretical spectrum with sensitivity curves planned for GW observations. To the author's knowledge, this methodology is new and never investigated in the literature before. In what follows, we discuss our main results.

\begin{figure}
\begin{center}
\includegraphics[width=3.3in]{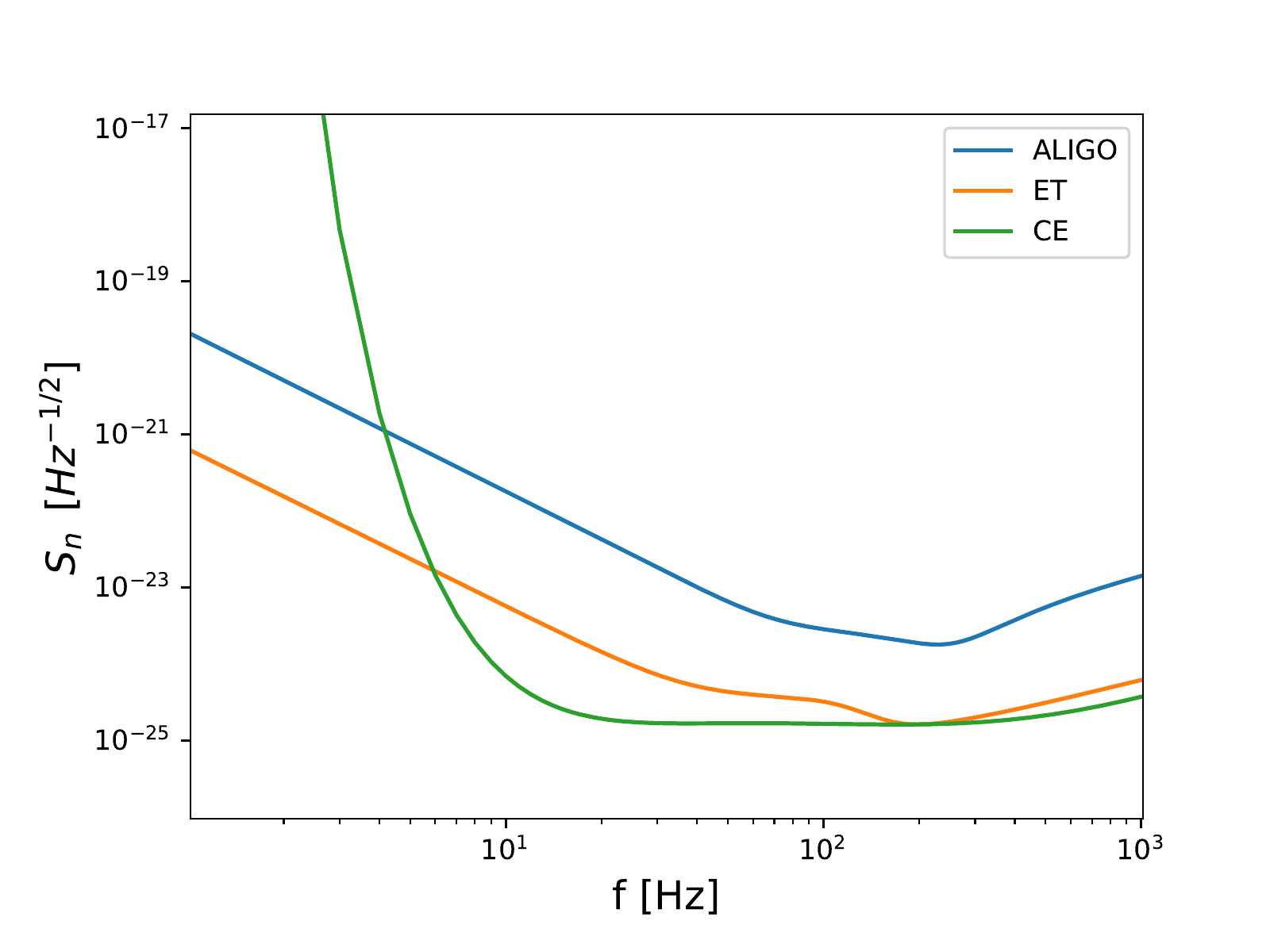}
\caption{Detector spectral noise density for the
Advanced LIGO (aLIGO), Einstein Telescope (ET) and Cosmic Explore (CE).}
\label{fig:S_n}
\end{center}
\end{figure}

\section{Results and Discussions}
\label{results}

First, we need to define some input properties of an AGWB, before performing numerical simulations. We list here the main ones.
\\

1 - We define $e(z) = \rho_{*}(z)/\rho_{*}(0)$, where $\rho_{*}(z)$ is the cosmic star formation rate density (in units of $M_\odot {\rm yr}^{-1} {\rm Mpc}^{-3}$). We consider $\rho_{*}(z)$ derived from the observations in \cite{star_formation}.
\\

2 - With respect to information about compact binary coalescences populations, we consider that $\Omega_{\rm GW}(f)$ as the sum due to contributions from stellar mass BBH, BNS and BBH-NS, i.e., $\Omega_{\rm GW}(f) = \Omega^{\rm BBH}_{\rm GW}(f) + \Omega^{\rm BNS}_{\rm GW}(f) + \Omega^{\rm BBH-NS}_{\rm GW}(f)$. These are the most promising GWs sources for ground-based interferometers. In this sense, we use $r_0$ values corresponding to the realistic estimates, $r_0 = 1, 0.03, 0.05$ ${\rm Mpc}^{-3} {\rm Myr}^{-1}$, for BNS, BBH-NS, BBH, respectively. Also, we replaced $M^{5/3}_c$ with $ \langle M^{5/3}_c \rangle$ to account for a distribution of system masses with an average over individual energy spectra as presented in \cite{AGWB1}.
\\

\begin{figure*}
\begin{center}
\includegraphics[width=3.in]{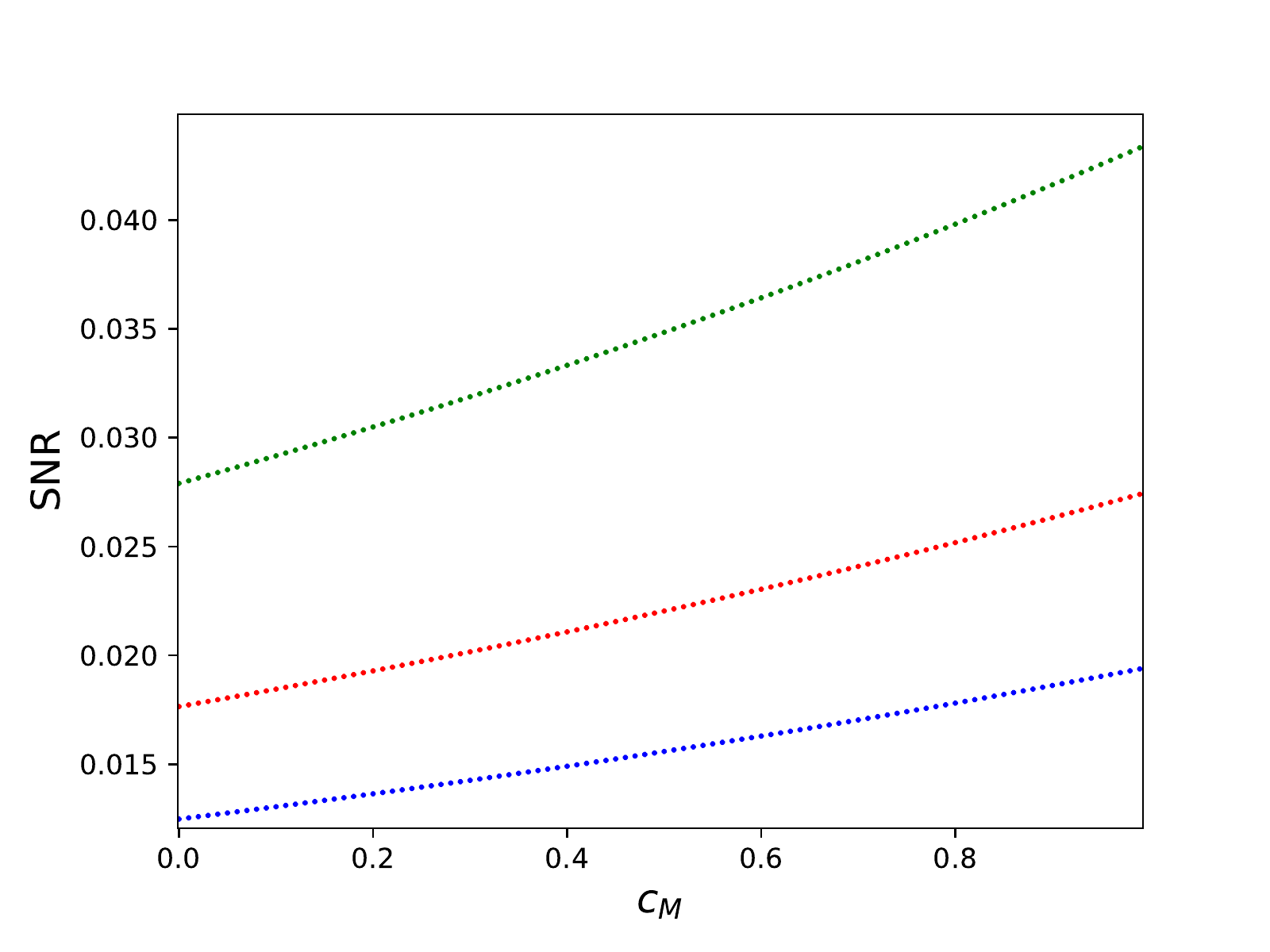} \,\,\,\,
\includegraphics[width=3.in]{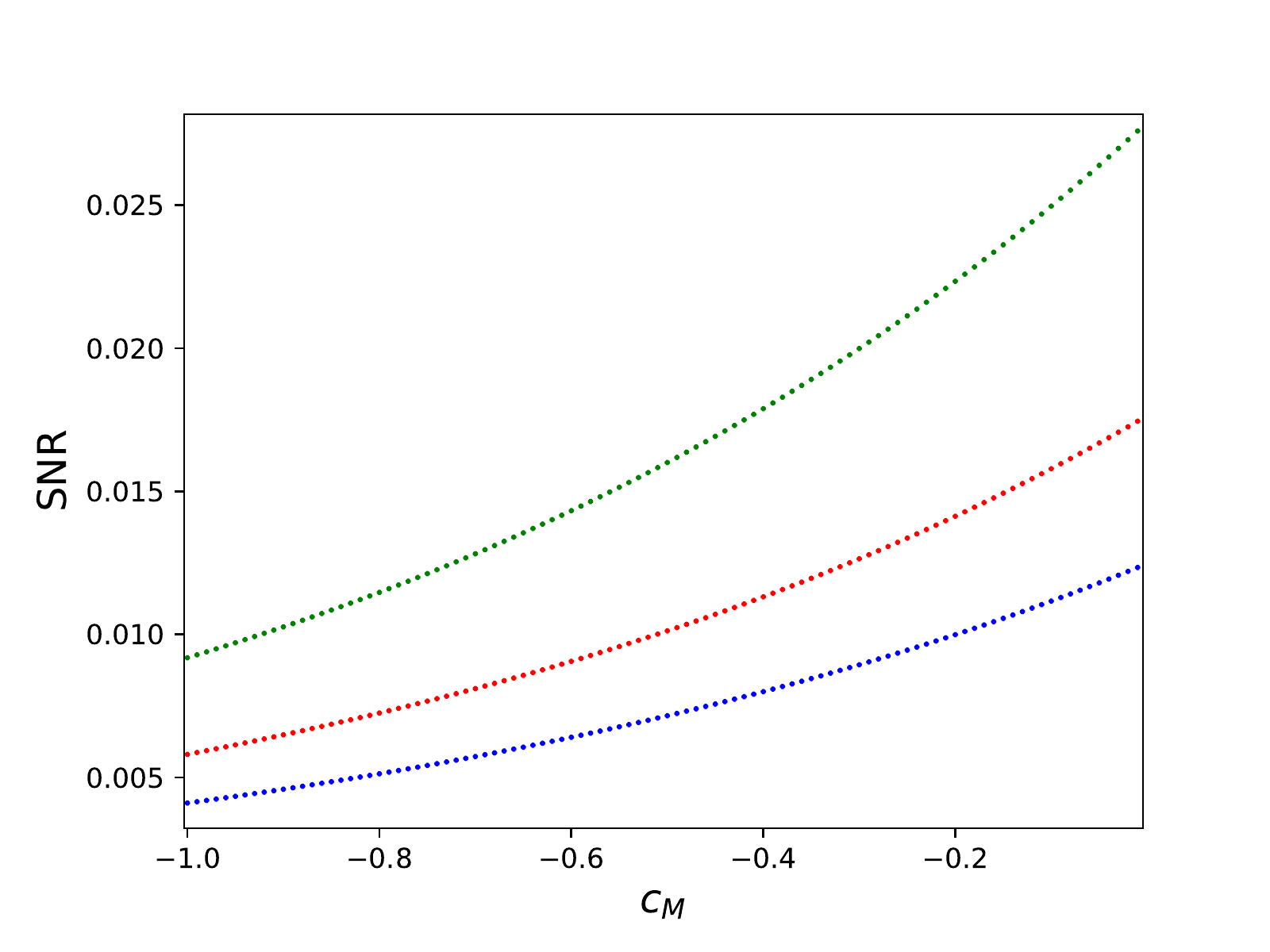}
\caption{The SNR as a function of the parameter $c_M$ (amplitude of the running of the Planck mass), assuming {\it Advanced LIGO} noise power spectral density sensibility for different observation times, 1 yr, 3 yr, and 5yr in blue, red and green, respectively. \textbf{Left panel}: Theoretical framework given by the scenario I. \textbf{Right panel}: Scenario II.}
\label{SNR_aligo}
\end{center}
\end{figure*}

\begin{figure*}
\begin{center}
\includegraphics[width=3.in]{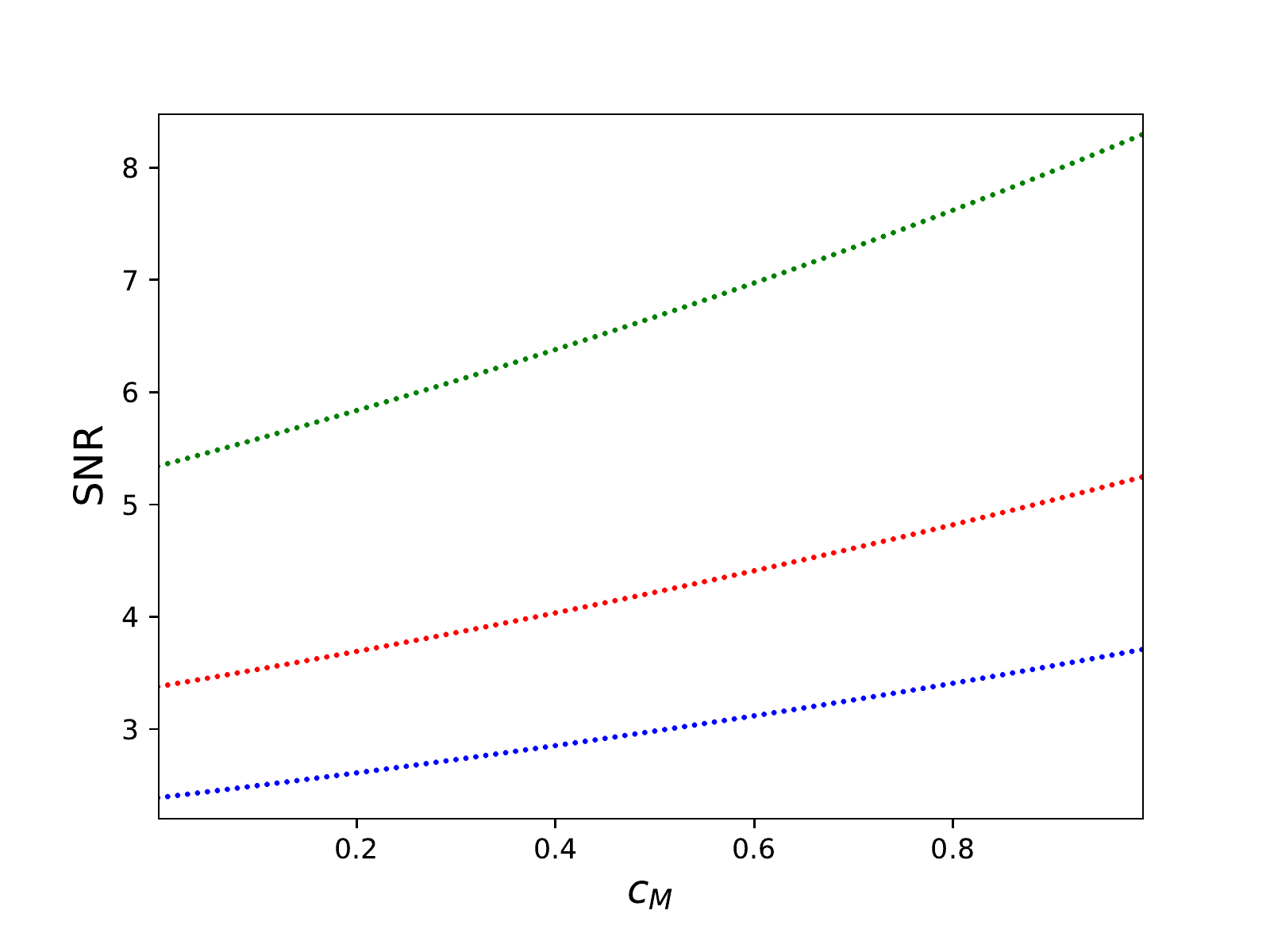} \,\,\,\,
\includegraphics[width=3.in]{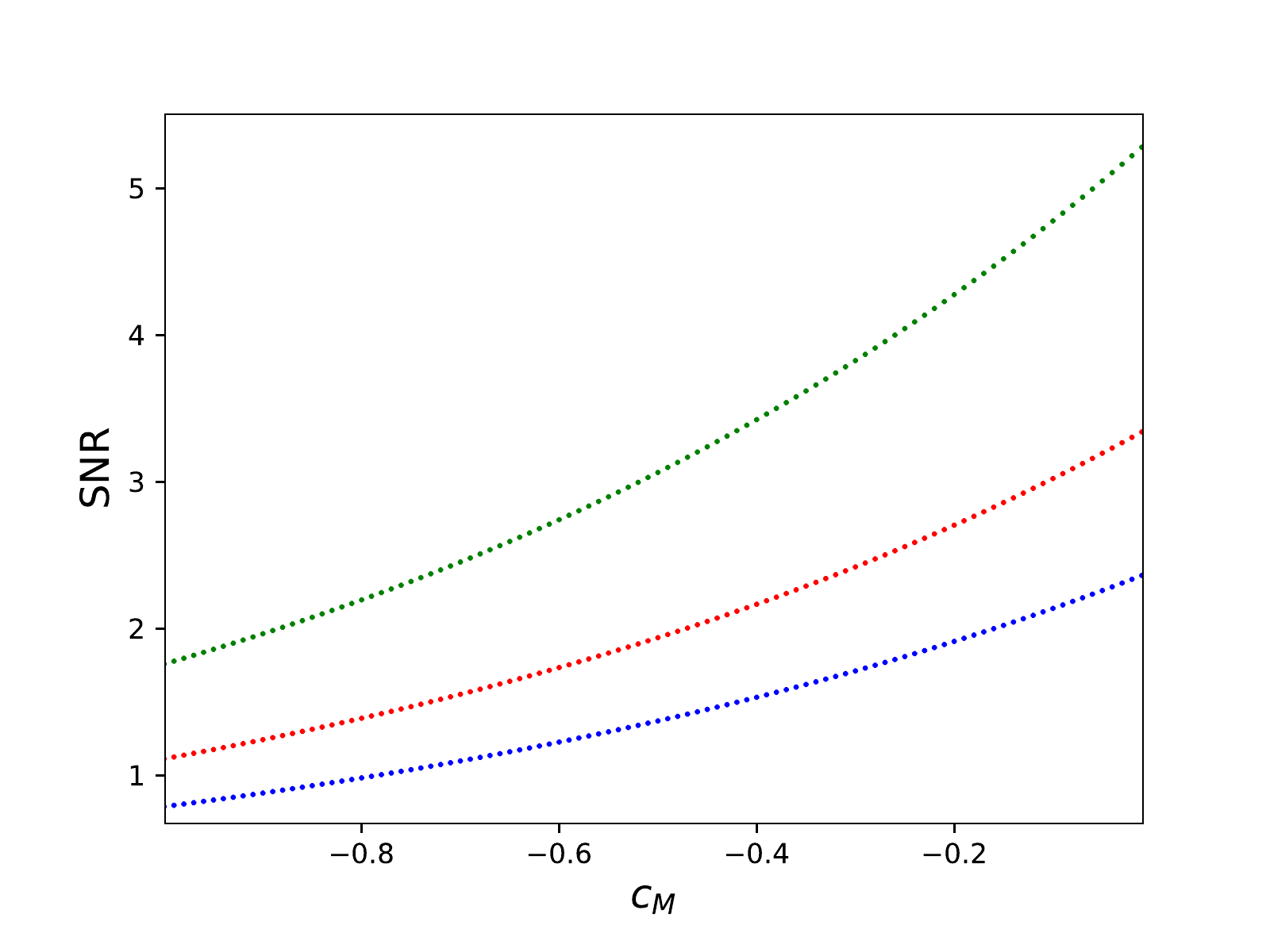}
\caption{The same as in Figs \ref{SNR_aligo}, but assuming {\it Einstein Telescope} noise power spectral density sensibility.}
\label{SNR_et}
\end{center}
\end{figure*}

\begin{figure*}
\begin{center}
\includegraphics[width=3.in]{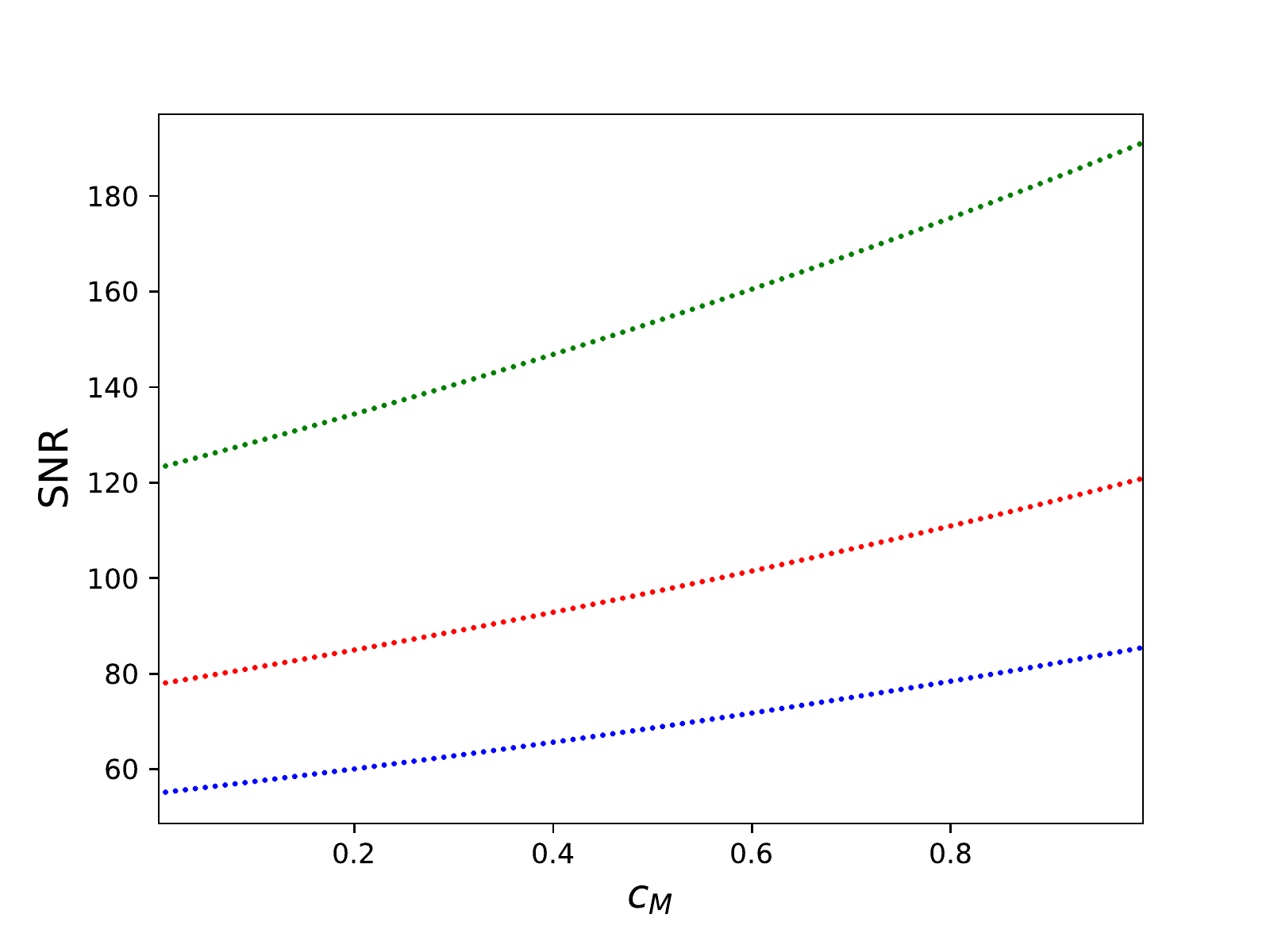} \,\,\,\,
\includegraphics[width=3.in]{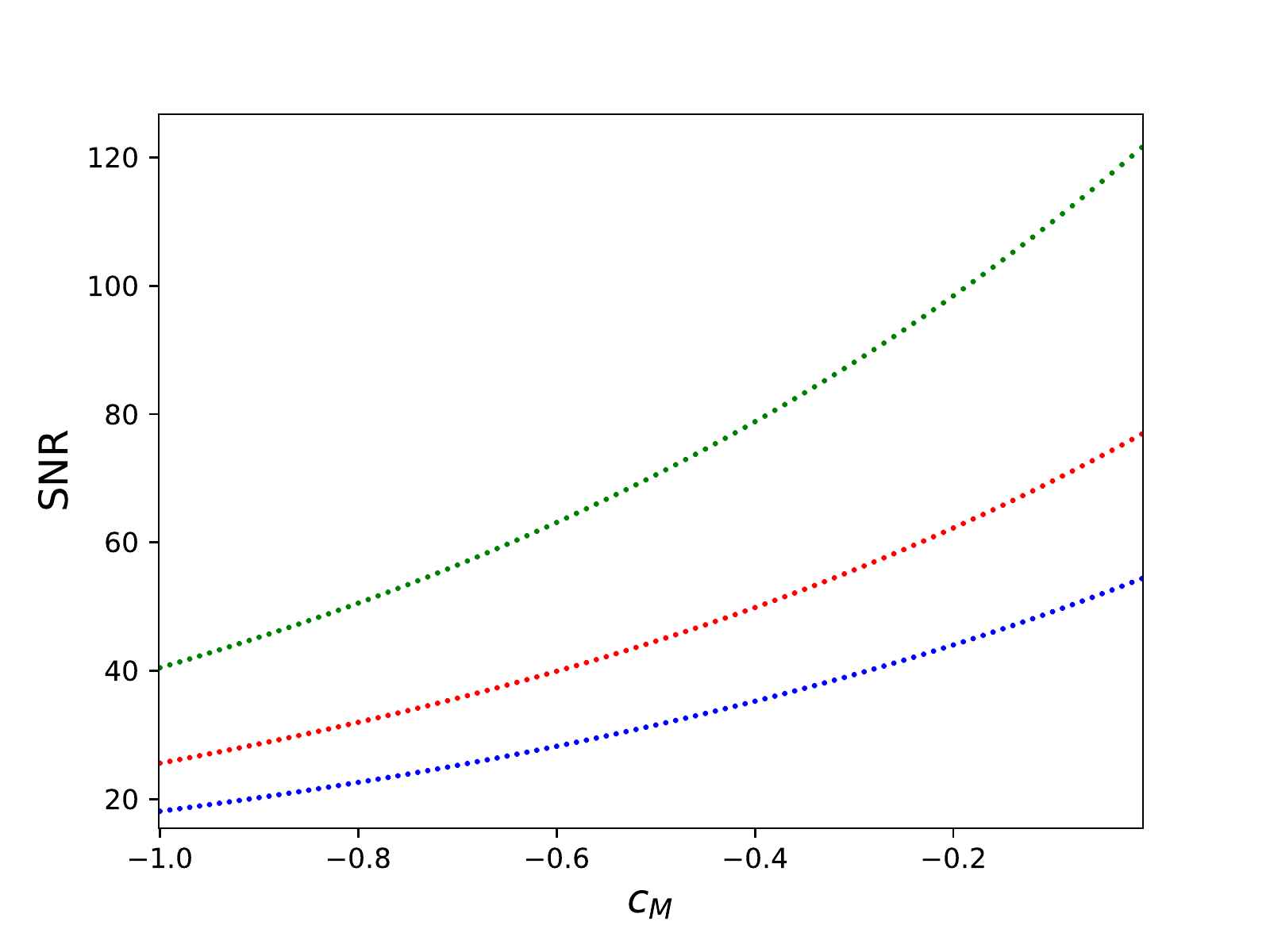}
\caption{The same as in Figs \ref{SNR_aligo} and \ref{SNR_et}, but assuming {\it Cosmic Explore} noise power spectral density sensibility. }
\label{SNR_ce}
\end{center}
\end{figure*}

From this input information, we can search our template applied to ground-based interferometers. We will consider the sensitivity as predicted by the detector spectral noise density for the Advanced LIGO (aLIGO) \cite{aLIGO}, Einstein Telescope (ET) \cite{ET} and Cosmic Explore (CE) \cite{CE}. Figure \ref{fig:S_n} shows $S_n$ for these detectors. The signal-to-noise ratio (SNR) to a SGWB in terms of the above quantities is

\begin{equation}
{\rm SNR} = \sqrt{T \int_{f_{\rm min}}^{f_{\rm max}} \Big(\frac{S_{h}}{S_{n}} \Big)^2}, 
\end{equation}
where $f_{\rm min}$, $f_{\rm max}$ denote, respectively, the minimal and maximal frequencies accessible at the detector and/or some range the interest for research. The SNR increases as the square root of the observation time $T$. In the above equation, we define \cite{Maggiore},

\begin{equation}
\label{Linder_model1}
S_{h}(f) = \frac{3H_0^2}{2 \pi^2} f^{-3} \Omega_{\rm GW}(f)
\end{equation}

Note that assuming an isotropic GW background, a factor of 1/5 should be included to account for the average detector response over all source locations in the sky. In what follows, in all results, we consider the input: $z_{\rm max} = 2$, $f_{\rm min} = 1$ Hz and $f_{\rm max} = 100$ Hz. These values are justified because up to $z = 2$, it comprises the range for the majority compact binary coalescences populations from stellar mass, which in return presents greater amplitude in the range $f \sim [1, 100]$. We use $H_0 =$ 70 km/s/Mpc and $\Omega_m= 0.31$ to fix the background expansion. 
Evaluation for values greater than this range, does not change the main results considerably.

Figure \ref{SNR_aligo} shows the SNR as a function of the parameter $c_M$ for both scenarios, and for different observation time periods within the sensibility of the aLIGO.  In all cases, the GR theory corresponds to $c_M = 0$. As expected, we have a signal and SNR very low, making its detectability difficult, for a wide range of intervals in $c_M$.

We can note a pattern around the features of the theoretical framework. For $\alpha_M > 0$, the SNR tends to increase, when $c_M$ also increases. For $\alpha_M < 0$, we notice the opposite, where $c_M$ decreases, we have SNR also decreasing. Once the $\alpha_M$ function quantifies a general property for all gravitational theories, we can conjecture that this should happen for any model beyond the GR in general, which can be written in terms of the $\alpha_M$.

Figures \ref{SNR_et} and \ref{SNR_ce} show the influence of the $c_M$ on the SNR from the spectral noise density for the ET and CE, respectively. Here, we also take different observation time periods. For ET, we find that only for $\gtrsim$ 5 yrs, we can note significant SNR values, which we can talk about for a possible observation of the signal for a AGWB. Note that GR always has SNR value smaller (larger), with respect to $c_M > 0 \,\, (< 0)$, respectively. Within CE sensitivity, we can detect a strong signal, with high SNR values for both scenarios, even for 1 yr of operation. This is because CE can be more sensitive than ET by up to 2 orders of magnitude in the band [1-100] Hz.

It is interesting to note that the residual foreground in the range [1- 100] Hz, should be considered in future ground-based stochastic searches for AGWB signal. In principle, this signal can be detected in GR as well as in modified gravity models, as shown in figures \ref{SNR_et} and \ref{SNR_ce}.

\begin{figure*}
\begin{center}
\includegraphics[width=3.in]{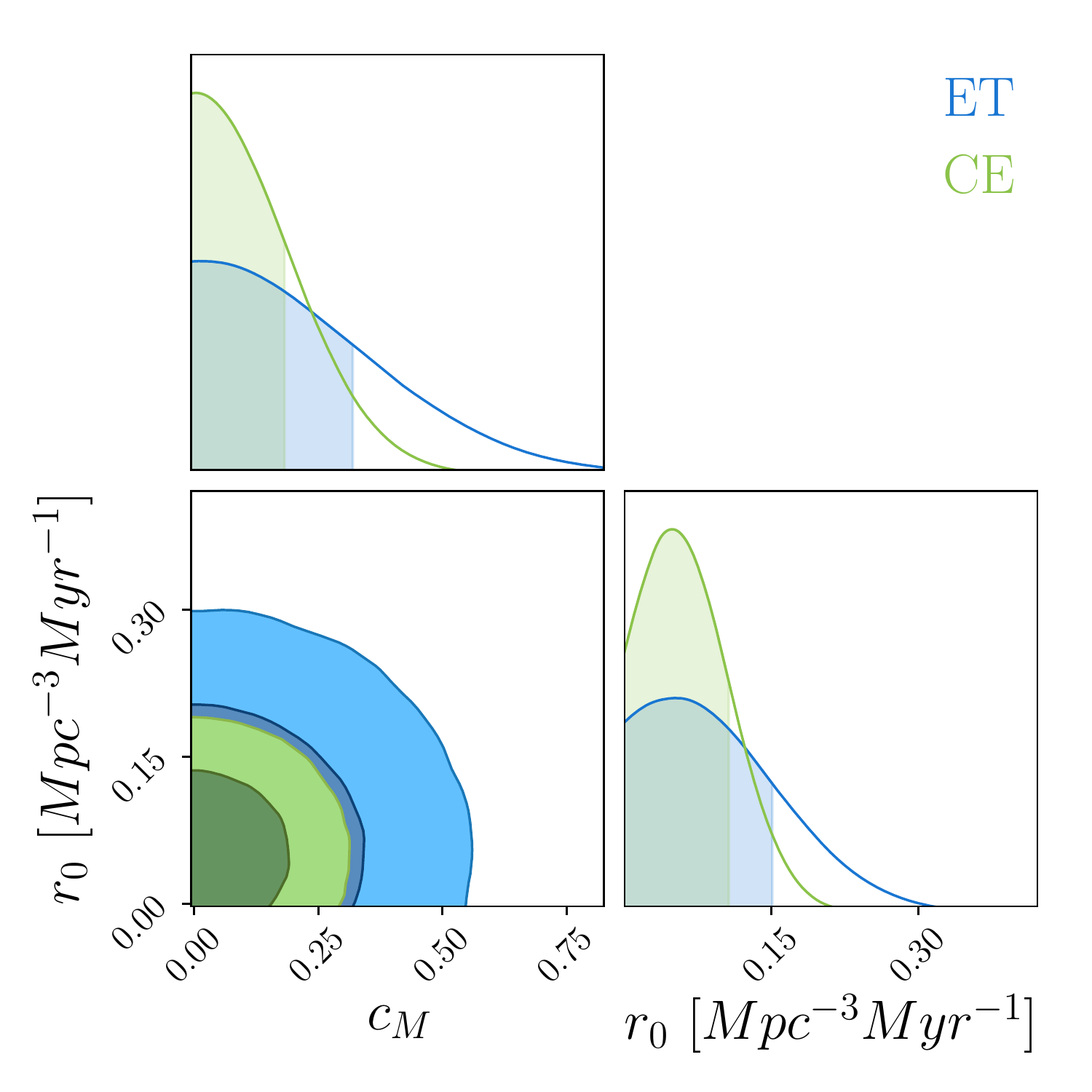} \,\,\,\,
\includegraphics[width=3.in]{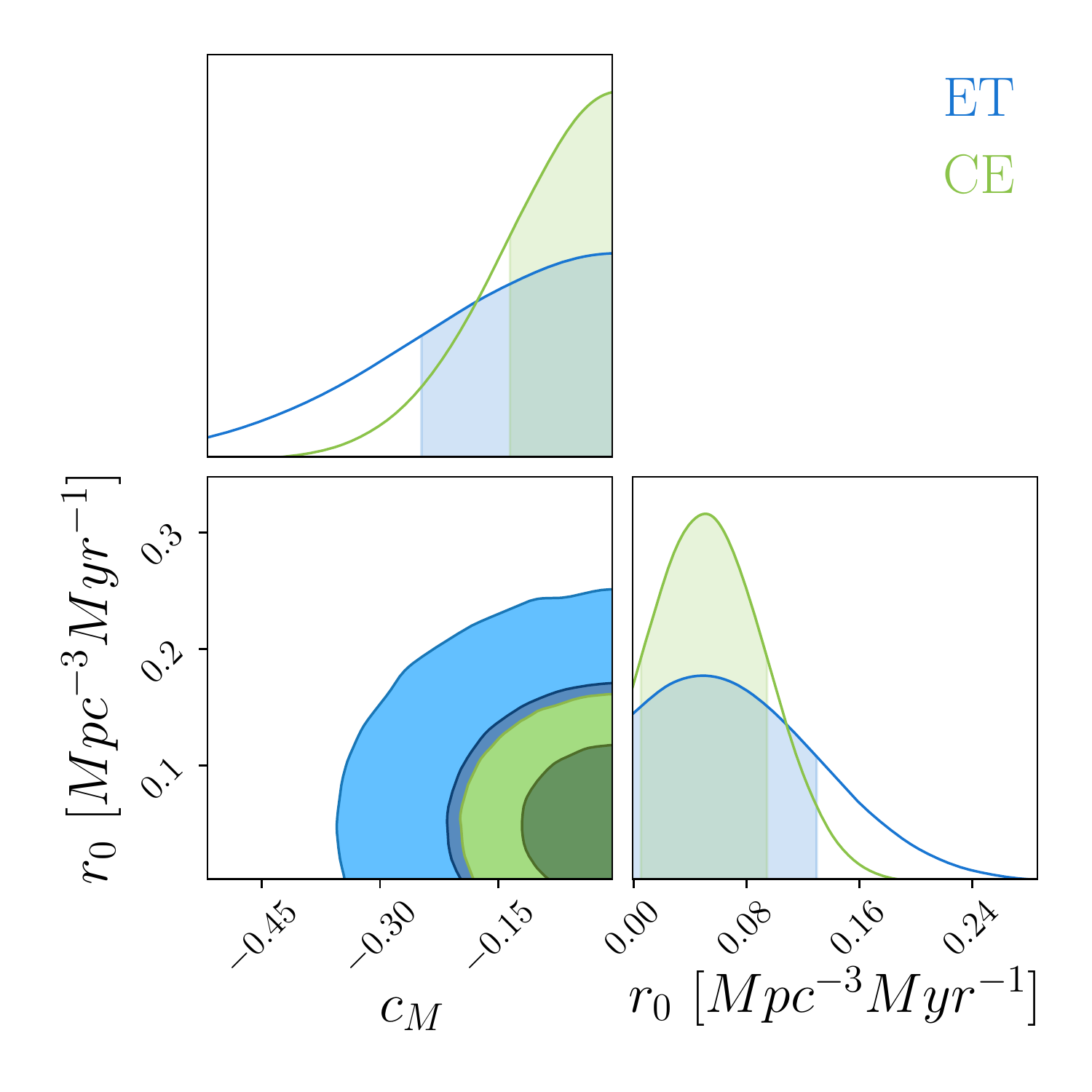}
\caption{Two-dimensional marginalized distributions of the free parameters $c_M$ (amplitude of the running of the Planck mass) and $r_0$ (local BBH rate density) forecasted from the ET and CE sensitivity. \textbf{Left panel}: Theoretical framework given by the scenario I. \textbf{Right panel}: Scenario II.}
\label{fig:Fisher_model}
\end{center}
\end{figure*}

For a high enough SNR, we can use the Fisher matrix analysis to provide upper bounds for the free parameters of the models. We refer the reader to \cite{Fisher01, Fisher02, Fisher03} for a discussion on the Fisher analysis in GWs signal. Thus, let us limit ourselves to apply the Fisher information only to ET and CE. The root-mean-squared error on any parameter is determined by

\begin{align}
\label{}
\Delta \theta^i = \sqrt{\Sigma^{ii}},
\end{align}
where $\Sigma^{ij}$ is the covariance matrix, i.e, the inverse of the Fisher matrix, $\Sigma^{ij} = \Gamma_{ij}^{-1}$. The Fisher matrix is given by

\begin{equation}
\label{}
\Gamma_{ij} = \left( \dfrac{\partial \tilde{h}}{\partial \theta^i} \mid  \dfrac{\partial \tilde{h}}{\partial \theta^j}  \right),
\end{equation}
where we define the inner product as

\begin{equation}
\label{}
(\tilde{h}_1 \mid \tilde{h}_2) \equiv 2 \int_{f_{low}}^{f_{upper}} \dfrac{\tilde{h}_1 \tilde{h}_2^{*} + \tilde{h}_1^{*} \tilde{h}_2}{S_n(f)} df,
\end{equation}
where the 'star' stands for complex conjugation, and $S_n(f)$ is the detector spectral noise density. The characteristic amplitude of a AGWB signal is given by $h= f S_h(f)$ \cite{Maggiore}.

Analyzing the eq. \ref{GW_4}, we can see that even within GR theory prediction, possible different values of the parameter $r_0$ can induce corrections in amplitude. Thus, when considering a parameter estimation in modified gravity context, $r_0$ estimation can play a important role, since this parameter can generate effects that can falsify possible real deviations due $\alpha_M$ contribution. We are assuming that the merger rate follows the star formation rate. So, greater uncertainty comes from $r_0$ for BBH. As shown in \cite{Fishbach}, the BH redshift distribution can lead to uncertainties, which in principle, should affect parameter estimation analysis. Thus, when analyzing forecasts on $c_M$, also let us take $r_0$ for BBH as a baseline parameter. We keep $r_0$ to be 1 and 0.03 ${\rm Mpc}^{-3} {\rm Myr}^{-1}$, for BNS and BBH-NS, respectively, once that these systems can be well modeled from star formation rate. Therefore, we can interpret the results below leading to an optimistic scenario.

Figure \ref{fig:Fisher_model} shows the marginalized distributions for the parameter $c_M$ and $r_0$ from the ET and CE sensitivity for both scenarios. 
For ET forecast, we find $c_M < 0.27$ and $c_M > -0.20$, at 1$\sigma$ confidence level (CL) in the framework of the scenarios I and II, respectively.
For CE forecast, we find  $c_M  < 0.18$ and $c_M  > -012$ at 1$\sigma$ CL, for the scenario I and II, respectively.

We can note that these bounds are of the accuracy matching the current measures \cite{Horndeski_constraints_01,Horndeski_constraints_02,Horndeski_constraints_03}. Thus, we can expect that future constraints using real data from a possible AGWB in this band, will impose strong limits on possible deviations from general relativity. 

\section{Final remarks}
\label{final}

The detectability of an isotropic and stochastic AGWB in the near future could open new ways to investigate fundamental physics, once many different astrophysical sources and physical properties contribute to the AGWB. The observational landscape is growing, and it covers a large range of frequencies where the AGWB is present. With regard to gravity, two main means can modify the AGWB properties. i) Generation mechanism of the signal by the sources ii) Modified propagation of the signal over cosmological volumes. In this work, we have investigated the latter in a parametric scenario given in terms of the Horndeski gravity. We find that gravitational-wave detector of third-generation like ET and CE, can detect the AGWB with significant SNR in our simple approximation, in particular from the perspectives of the CE experiment, in both, GR theory and modified gravity. Within the sensitivity of these instruments, a forecast analysis shows that the corrections on the amplitude of the running of the Planck mass can be bounded with the same precision as current measures.
Therefore, an AGWB signal can put significant  bounds in modified gravity models, and certainly some tight constraints in combination with other data sets.

Others interesting astrophysical sources can contribute to the AGWB in frequency well below the 1 Hz like supermassive black hole, primordial black hole, binaries white dwarfs, r-mode instability of neutron stars, as well as several phenomenological physical aspects in the early Universe. It may be interesting to investigate possible corrections on this type of signal, which is in LISA band frequency, as well as to consider a network of interferometric detectors like LISA + ET and/or LISA+CE, and explore the parameter space of some models to probe gravity bounds.

\begin{acknowledgments}
\noindent 
I thank the referee for some clarifying points. I am grateful to Jos\'e C. N. de Araujo, Michele Maggiore, Suresh Kumar and Eric Linder for very constructive comments and suggestions. Also, I would like to thank the agency FAPESP for financial support under the project No. 2018/18036-5 and thematic project 2013/26258-4.
\end{acknowledgments}

\section*{Appendix: The Horndeski gravity}
\label{Horndeski_gravity}

In this appendix, we briefly review the functions in Horndeski
gravity, used in the main text of the paper. The
Horndeski theories of gravity \cite{Horndeski,Deffayet,Kobayashi11} are the most general Lorentz invariant scalar-tensor theories with second-order equations of motion. The Horndeski action reads

\begin{equation}
\label{acao_geral}
 S =  \int d^4 x \sqrt{-g} \Big[ \sum_{i=2}^{5} \frac{1}{8 \pi G} \mathcal{L}_i + \mathcal{L}_m \Big],
\end{equation}

\begin{equation}
 \mathcal{L}_2 = G_2(\phi, X),
\end{equation}

\begin{equation}
 \mathcal{L}_3 = -G_3(\phi, X) \Box \phi,
\end{equation}

\begin{equation}
 \mathcal{L}_4 = -G_4(\phi, X)R + G_{4X} [( \Box \phi)^2 - \phi_{;\mu \nu} \phi^{;\mu \nu}],
\end{equation}

\begin{align}
\mathcal{L}_5 = -G_5(\phi, X)G_{\mu \nu}\phi^{;\mu \nu} -
\frac{1}{6}G_{5X}[(\Box \phi)^3  + \\
2 \phi_{;\mu \nu}  \phi^{;\mu \sigma} \phi^{;\nu}_{;\sigma}
 - 3 \phi_{;\mu \nu} \phi^{;\mu \nu} \Box \phi],
\end{align}
where the functions $G_i$ ($i$ runs over 2, 3, 4, 5) depend on $\phi$ and $X = -1/2 \nabla^\nu \phi \nabla_\nu \phi $, with $G_{i X} = \partial G_i/\partial X$. For $G_2 = \Lambda$, $G_4 = M^2_p/2$ and $G_3 = G_5  = 0$, we recover GR with a cosmological constant. 
The running of the Planck mass, $\alpha_M$, is given by \cite{Bellini_Sawicki}

\begin{equation}
\label{alphaM}
\alpha_M = \frac{1}{H M^2_{*}} \frac{dM^2_{*}}{dt},
\end{equation}
where 

\begin{equation}
\label{A2}
M^2_{*} = 2(G_4 - 2XG_{4X} + XG_{5 \phi} - \dot{\phi} H X G_{5X}),
\end{equation}
is the effective Planck mass.

Other relevant quantity for the GWs context is the tensor speed excess, $\alpha_T$, which can be written as \cite{Bellini_Sawicki}

\begin{equation}
\label{alphaT}
\alpha_T = \frac{2 X (2G_{4X} - 2G_{5 \phi} - (\ddot{\phi} - \dot{\phi} H)G_{5X})}{M^2_{*}}.
\end{equation}

The functions $\alpha_M$ and $\alpha_T$ depend on the parameters of the theory and on the cosmological dynamics of the scalar field. On the other hand, the event GW170817 showed that the speed of GW, $c_T$, is very close to that of light for $z < 0.01$, that is, $ |c_T/c - 1| < 10^{-15}$. Thus, we have $\alpha_{T0} \simeq 0$, leading to consider $G_{4X} = G_{5 \phi} = G_{5X} \simeq 0$. Thus, under that consideration, from eq. (\ref{A2}), we can write 

\begin{equation}
M^2_{*} = 2 G_4.
\end{equation}

\end{document}